\documentstyle[aps,12pt,preprint,psfig]{revtex}
\tightenlines

\def \to {\rightarrow}
\def \beq {\begin{equation}}
\def \eeq {\end{equation}}
\def \ba {\begin{eqnarray}}
\def \ea {\end{eqnarray}}
\def \jpsi {J/\psi}

\def \< {\left <}
\def \> {\right >}
\def \os {\< {\cal O}_8^\psi({}^3S_1)\> }

\begin{document}
\preprint{
\hbox{PKU-TP-98-21\\
        hep-ph/9804295}}
\draft

\title{Diffractive $\jpsi$ production through Color-Octet mechanism at
        hadron colliders}

\author{Feng Yuan, Jia-Sheng Xu,  Hong-An Peng}
\address{\small {\it Department of Physics, Peking University, Beijing 100871, People's Republic
of China\\
         and China Center of Advanced Science and Technology (World Laboratory), Beijing 100080,
        People's Republic of China}}
\author{Kuang-Ta Chao}
\address{\small {\it China Center of Advanced Science and Technology (World Laboratory), Beijing 100080,
        People's Republic of China\\
      and Department of Physics, Peking University, Beijing 100871, People's Republic of China}}

\maketitle
\begin{abstract}

    We propose the color-octet mechanism combined with the two gluon
exchange model for the diffractive $J/\psi$ production
in hadron collisions. In the leading logarithmic approximation (LLA)
in QCD, we find that the diffractive $J/\psi$ production rate
is related to the off-diagonal gluon density 
in the proton and to
the nonperturbative color-octet matrix element of $J/\psi$. 
The rate is found to be very sensitive 
to the gluon density at very small values of $x$ (down to $x=O(10^{-6})$). 
As a result, this process may provide a wide window
for testing the two-gluon exchange model, and may be particularly 
useful in studying the small $x$ physics.
And it may also be a golden place to test the color-octet mechanism
proposed by solving the $\psi'(J/\psi)$ surplus problem at the Tevatron.

\end{abstract}


In recent years, there has been a renaissance of interest in
diffractive scattering.
These diffractive processes are described by the Regge theory in
terms of the Pomeron ($I\!\! P$) exchange\cite{pomeron}.
The Pomeron carries quantum numbers of the vacuum, so it is a colorless entity
in QCD language, which may lead to the ``rapidity gap" events in experiments.
However, the nature of the Pomeron and its interaction with hadrons remain a mystery.
For a long time it had been understood that the dynamics of the
``soft Pomeron'' was deeply tied to confinement.
However, it has been realized now that much can be learnt about
QCD from the wide variety of small-$x$ and hard diffractive processes,
which are now under study experimentally.
Of all these processes, the diffractive heavy quarkonium production
has drawn specially attention, because their large masses provide a natural
scale to guarantee the application of perturbative QCD.
In Refs.\cite{th1,th2}, the diffractive $\jpsi$ and $\Upsilon$ production
cross sections have been formulated in photoproduction processes and in
DIS processes in perturbative QCD.
In the framework of perturbative QCD the Pomeron is assumed to be
represented by a pair of gluons in the color-singlet state.
This two-gluon exchange model can successfully describe the experimental
results from HERA\cite{hera-ex}.
An important feature of this perturbative QCD model prediction is that
the cross section for the diffractive $\jpsi$ production is expressed
in terms of the square of the gluon density.

So far the previous studies are focused on the diffractive processes
at $ep$ collider (photoproduction and DIS processes),
we should expect that the two-gluon exchange model can also be used
to describe the diffractive processes at hadron colliders.
In this paper, we extend the idea of perturbative QCD description of
diffractive processes from $ep$ colliders to hadron colliders.
Unlike in the case of the photoproduction process,
at hadron colliders (shown in Fig.1), in the diffractive
process the final state $c\bar c$ pair, which carry the quantum number
of the incident gluon, is in a color-octet configuration.
This color-octet $c \bar c$ pair must evolve into the physical
color-singlet bound state of $\jpsi$ for the experimental observation.
Here we use the color-octet mechanism based on the NRQCD factorization
formalism\cite{nrqcd} to describe this evolving process.
In the last few years, the application of the color-octet mechanism has gained
some important success.
For example, by including this color-octet mechanism one might explain the
$\psi^\prime$ ($\jpsi$) surplus
measured by the CDF Collaboration at the Tevatron\cite{surplus}.

The diffractive $\jpsi$ production process in Fig.1 can be realized via
the color-octet ${}^3S_1$ channel, in which the $c\bar c$ pair in a
configuration of ${}^3S_1^{(8)}$ is produced in hard process as the
incident gluon interacts with the proton by $t$ channel color-singlet
exchange (the two-gluon ladder parametrized Pomeron),
and then evolve into the physical state $\jpsi$ through the long distance
process (nonperturbative).
In this long distance process some soft gluons will be emitted to change
the color of the $c\bar c$ pair to coincide with the quantum number of the
final state $\jpsi$.
The soft gluons carry little momentum and will cause little change to the
final state $\jpsi$ spectrum in the diffractive processes.
In the old color-singlet model\cite{csm} the emitted gluons
are hard, which will lead to a quite different momentum distribution in
experiment.
In NRQCD, the long distance evolving process is described by the
nonperturbative matrix elements of four-fermion operators.
In the process we considered in this paper,
the associated matrix element is $\< {\cal O}_8^\psi({}^3S_1)\> $,
which is also used to describe the large $p_t$ prompt $\jpsi$
production at the Tevatron and has been determined by several authors\cite{surplus}.

Based on the NRQCD factorization formalism, as a general factorization ansatz,
the rate for the diffractive
process $g p\to \psi p$\footnote{
Exactly, the final state of this diffractive process should include
some soft gluons.
However, becuse the soft gluons only carry little momenta, they do not
change the final state kinematic distribution much.
In this paper, we neglect the soft gluon effects in the long distance
evolving process.}
can be factorized into the short distance part and the long distance part as
the following form,
\beq
\label{e1}
|{\cal A}(g p\to J/\psi p)|^2=|{\cal A}(g p\to (c\bar c)[{}^3S_1^{(8)}]p)|^2
        \times |{\cal A}((c\bar c)[{}^3S_1^{(8)}]\to J/\psi)|^2.
\eeq
Here ${\cal A}(g p\to (c\bar c)[{}^3S_1^{(8)}]p)$
is the short distance amplitude which 
can be calculated in perturbative QCD. The amplitude 
${\cal A}((c\bar c)[{}^3S_1^{(8)}]\to J/\psi)$ describes the
long distance evolving process which is related to the 
nonperturbative color-octet matrix element as
$|{\cal A}((c\bar c)[{}^3S_1^{(8)}]\to J/\psi)|^2=
        \< {\cal O}_8^\psi({}^3S_1)\> /{24m_c}$\cite{nrqcd}.

For the diffractive subprocesses, $gp\to J/\psi p$, the leading contribution
comes from the diagrams shown in Fig.2.
Due to the positive signature of these diagrams (color-singlet state),
we know that the real part of the amplitude cancels out in the leading
logarithmic approximation.
The first two diagrams are similar to those calculated in photoproduction
process, and the rest diagrams are new due to the existence of the
gluon-gluon interaction vertex.
From the following calculations, we can see that the diagrams other than the
first two are needed to guarantee the gauge invariance.

For convenience, we perform our calculations in terms of the Sudakov variables.
That is, for the involved particles the four momenta are decomposed as,
$k_i=\alpha_i q+\beta_i p+\vec{k}_{iT}$,
where $q$ and $p$ are the momenta of the incident gluon and the proton,
$q^2=0$, $p^2=0$, and $2p\cdot q=W^2=s$.
Here $s$ is the total c.m. energy of the gluon-proton system,
$\alpha_i$ and $\beta_i$ are the momentum fraction of $q$ and $p$ carried by
the momentum $k$.
$k_{iT}$ is the transverse momentum, which satisfies $k_{iT}\cdot q=0$,
$k_{iT}\cdot p=0$.
In our calculations, we set the momentum transfer $t$ equal to zero,
i.e., $t=(k-k^\prime)^2=0$.
All of these Sudakov variables are fixed by considering the mass shell
condition on the crossed lines shown in Fig.2.
Take Fig.2(a) as an example, the Sudakov variables are,
$$\alpha_k=\alpha_{k^\prime}=-\frac{k_T^2}{s},~~
\beta_k=-\frac{M_\psi^2+2k_T^2}{s},~~\beta_{k^\prime}=-\frac{2k_T^2}{s},$$
where $M_\psi$ is the mass of $\jpsi$ which is equal to $2m_c$.

Following Ref.\cite{th1}, the calculations are straightforward.
When we perform the integral over the loop momentum $k$, the main large
logarithmic contribution comes from the region
${1\over R_N^2}\ll k_T^2\ll M_\psi^2$ ($R_N$ is the nucleon radius)\cite{th1}.
So, we calculate the amplitude as an expansion
of $k_T^2$.

For Fig.2(a), the imaginary part of the short distance amplitude
${\cal A}(g_a p\to (c\bar c)_b[{}^3S_1^{(8)}]p)$ is,
to leading order contribution,
\beq
\label{ima}
{\rm Im}{\cal A}^{(a)}=F\times
       {{1\over 9}\delta^{ab}}\int\frac{dk_T^2}{k_T^4}\frac{1}{m_c^2}G(k),
\eeq
where $F=\frac{3\pi}{2}g_s^3m_cs$,
$a$ and $b$ are the color indexes of the incident gluon and
the $c\bar c$ pair in color-octet ${}^3S_1$ state.
Factor $1\over 9$ is the color factor.
The function $G(k)$ specifies the probability of finding the gluon in
the proton. In the simplest three valence quark model,
$G=\frac{4}{3}\frac{\alpha_s}{\pi}\times 3$.
For Fig.2(b), the result is,
\beq
\label{imb}
{\rm Im}{\cal A}^{(b)}=-F\times(
       {-{1\over 72}\delta^{ab}})\int\frac{dk_T^2}{k_T^4}\frac{1}{m_c^2+k_T^2}G(k),
\eeq
where the color factor is $- {1\over 72}$.
Unlike in the case of the diffractive photoproduction processes, the color
factors of these two diagrams (Fig.2(a) and Fig.2(b)) are not the same.
The leading part of the contributions from these two diagrams (which is
proportional to $1\over k_T^4$) can not cancel out each other.
After integrating the loop momentum $k$, for small $k_T$ this will lead to a linear
singularity, not a logarithmic singularity (proper in QCD) as that in
diffractive photoproduction process \cite{th1}.
So, here in the case of diffractive process at hadron colliders,
there must be some other diagrams to cancel out the leading part of
Fig.2(a) and Fig.2(b) to obtain the correct result.
This is also due to the gauge invariance requirement.
As mentioned above, in QCD due to the nonabelian $SU(3)$ gauge theory
there are additional diagrams shown in Fig.2(c)-(e) as compared
with that in photoproduction at $ep$ colliders.
By summing up all these diagrams together, we expect that in the final
result the leading part singularity which is
proportional to $1\over k_T^4$ will be canceled out,
and only the terms proportional to $1\over k_T^2$ will be retained.

The contribution from Fig.2(c) is,
\beq
\label{imc}
{\rm Im}{\cal A}^{(c)}=F\times
       ({-{1\over 2}\delta^{ab}})\int\frac{dk_T^2}{k_T^4}\frac{1}{4m_c^2+k_T^2}G(k),
\eeq
where $-{1\over 2}$ is color factor.
From Eqs.(\ref{ima}), (\ref{imb}), (\ref{imc}), we
can see that the leading part singularity from Fig.2(a)-(c) are
canceled out as expected.

By the same reason, for Fig.2(d) and Fig.2(e), the leading part of each
diagram is proportional to $1\over k_T^4$. However, their sum is only
proportional to $1\over k_T^2$ because the leading part is canceled out.
Their final results is,
\beq
{\rm Im}{\cal A}^{(de)}=F\times
       ({-{1\over 2}\delta^{ab}})\int\frac{dk_T^2}{k_T^2}\frac{2}{16m_c^4}G(k).
\eeq

Adding all the contributions from Fig.2(a)-(e), we get
the imaginary part of the short distance amplitude,
\beq
{\rm Im}{\cal A}(gp\to (c\bar c)[{}^3S_1^{(8)}]p)=F\times
       {(-{13\over 18}\delta^{ab})}\int\frac{dk_T^2}{k_T^2}\frac{1}{M_\psi^4}f(x',x^{\prime\prime};k_T^2).
\eeq
Here, we rewrite the function $G(k)$ as $G(k)=f(x',x^{\prime\prime};k_T^2)$,
\beq
f(x',x^{\prime\prime};k_T^2)=\frac{\partial G(x',x^{\prime\prime};k_T^2)}{\partial {\rm ln} k_T^2}
\eeq
where the function
$G(x',x^{\prime\prime};k_T^2)$ is the so-called
off-diagonal gluon distribution function\cite{offd}.
Here, $x'$ and $x^{\prime\prime}$ are the momentum fraction of the proton
carried by the two gluons.
It is expected that for small $x$, there is no big difference between the off-diagonal and
the usual diagonal gluon densities\cite{off-diag}.
So, in the following calculations, we estimate the production rate by
approximating the off-diagonal gluon density by 
the usual diagonal gluon density, 
$G(x',x^{\prime\prime};Q^2)\approx xg(x,Q^2)$, where $x=M_\psi^2/s$.

Finally, in the leading logarithmic approximation (LLA), we obtain
the imaginary part of the short distance amplitude,
\beq
\label{e3}
{\rm Im}{\cal A}(gp\to (c\bar c)[{}^3S_1^{(8)}]p)=
{(-{13\over 18}\delta^{ab})}\frac{F}{M_\psi^4}xg(x,\bar{Q}^2),
\eeq
where we set $\bar{Q}^2=M_\psi^2$.
The factorization scale in the gluon density is very important because
we know that parton distributions at small $x$ change rapidly with
this scale\cite{recent}.
In the literature there are various choices\cite{th1,th2,th3}.
Here, we choose the scale to be $M_\psi=2m_c$
which is typically used in NRQCD factorization approach\cite{nrqcd}.

Using Eqs.(\ref{e1}) and (\ref{e3}), we get the cross section
for the diffractive subprocess $gp\to \jpsi p$,
\beq
\label{e4}
\frac{d\hat\sigma(gp\to \jpsi p)}{dt}|_{t=0}=\frac{|{\cal A}|^2}{16\pi s^2}=
        \frac{169\pi^4m_c}{32\times 27}\frac{\alpha_s(\bar{Q}^2)^3\os }{M_\psi^8}
        [xg(x,\bar{Q}^2)]^2.        
\eeq

We have also calculated the color-octet ${}^1S_0$ and ${}^3P_J$ subprocesses.
As expected they do not contribute to the diffractive $\jpsi$ production
at hadron colliders.
That is, only the color-octet ${}^3S_1$ contributes in this
process. So the diffractive $\jpsi$ production considered in this paper
is sensitive to the matrix element $\< {\cal O}_8^\psi({}^3S_1)\> $, which
is very important to describe the prompt $\jpsi$ production at the
Tevatron\cite{surplus}.
Therefore, the diffractive $\jpsi$ production at hadron colliders
would provide a golden place to test the color-octet mechanism in the
heavy quarkonium production.

Provided the partonic cross section (\ref{e4}) above, we can get the cross
section of diffractive $\jpsi$ production at hadron level.
However, as we know, there exsit nonfactorization effects in the hard
diffractive processes at haron collisions\cite{preqcd,collins,soper,tev}.
As argued by D.E. Soper, these effects may be
accounted by a suppression factor ${\cal F}_S$.
At the Tevatron,
the value of ${\cal F}_S$ may be as small as ${\cal F}_S\approx 0.1$\cite{soper,tev}.
That is to say, the total cross section of the diffractive processes
at the Tevatron may be reduced down by an order of magnitude due to
the nonfactorization effects.
So, the cross section for the diffractive process $pp(\bar p)\to \jpsi p$
can be formulated as
\beq
\label{e5}
\frac{d\sigma(pp(\bar p)\to \jpsi p)}{dt}|_{t=0}
        ={\cal F}_S\frac{169\pi^4m_c}{32\times 27}\frac{\alpha_s(\bar{Q}^2)^3\os }{M_\psi^8}
        \int_{x_{1\rm min}} dx_1g(x_1,\bar Q^2)[xg(x,\bar{Q}^2)]^2,
\eeq
where $x_1$ is the longitude momentum fraction of the proton (or antiproton)
carried by the incident gluon.
So, the c.m. energy of the gluon-proton
system is $s=x_1S$, where $S$ is the total c.m. energy of the proton
and proton (antiproton) system (e.g., $S=(1800GeV)^2$ at the Tevatron).
Then, $x=M_\psi^2/s=M_\psi^2/x_1S$.
We can see that at the Tevatron, $x$ may lower down to the $10^{-6}$ level.
The lower bound of the integral variable $x_1$ is set to satisfy the relation 
$M_\psi^2\ll s$ which guarantee the validity of our calculations for this high
energy process.

For the numerical calculations, we choose the input parameters as,
$M_\psi=3.10GeV$, $\alpha_s(2m_c)=0.27$, $\os =0.0106 GeV^{3}$.
The value of the color-octet matrix element is taken from \cite{benek}.
For the parton distribution function, we use the GRV NLO set\cite{grv}.
The region for $x$ is limited in GRV set to $10^{-5}<x<1$, but in our case,
the value of $x$ can reach the level down to $10^{-6}$, so we
extrapolate the GRV set to these values according to their fitting
parameters.

In Fig.3, we plot the differential cross section $d\sigma/dt(t=0)$
as a function of $x_{1{\rm min}}$ for $p\bar p$ collider at the Fermilab Tevatron.
(The value of the suppresion factor of ${\cal F}_S$ is taken as $0.1$).
From this figure, we see that the main contribution to the total cross
section comes from the region of $x_1>10^{-3}$, which contributes $87\%$
of the total cross section.
$x_1>10^{-3}$ corresponds to the region of $x<2.9\times 10^{-3}$.
For $x_1>10^{-2}$, the contribution is over $58\%$, which corresponds
to the region of $x<2.9\times 10^{-4}$.
For $x_1> 10^{-1}$, the
contribution is above $16\%$, which corresponds to $x<2.9\times 10^{-5}$.
We can see that the gluon density at small $x$ is very important to
the diffractive $\jpsi$ production at hadron colliders.
Therefore, we may measure the gluon density at small $x$
by observe the diffractive $\jpsi$ production.
At the Tevatron, the gluon density can be probed down to the level of 
$10^{-6}$ through this process.
At the LHC, the situation is even more optimistic.

After integrating over $x_1$, we get the total cross section for the
diffractive production $\jpsi$ at the Tevatron,
$\sigma(p\bar p\to \jpsi p)=66 {\rm nb}.$
Here, we assume the $\jpsi$ diffractive slope $b$ to take the same value
as in the photoproduction process, i.e., $b=4.5GeV^{-2}$\cite{hera-ex}.
We have also estimated the gluon $k_T$ effects following
\cite{th1}, which will increase the total cross section by a
factor within $5\%$.
Our result shows that the rate of this process is more than three
order of magnitude larger than
that of large $p_T$ $\jpsi$ production
in diffractive process by soft Pomeron exchange\cite{lpt},
where the cross section at large $p_T$ ($\geq 8GeV$) is of
order of $0.01 {\rm nb}$.

However, the above results must be viewed as a rough estimate because in our
calculations there are large theoretical uncertainties.
The parton distribution function (PDF) is unknown to such low $x$ region,
and different sets of PDF will result in different result for the cross
section.
And the main uncertainty may come from the charm quark mass
because the cross section sales like the eighth power of $m_c$.
Even a modest change in the charm mass $m_c$ will result in huge
changes in the overall normalization of the cross section.
Another uncertainty may come from the color-octet matrix element
$\os $. In our estimate, we follow the determination of Ref.\cite{benek}.
So far our formula is only to the LLA level,
higher order corrections are neglected.
Nevertheless, as a first order approximation, our result may provide much sense
about the diffractive $\jpsi$ production rate at hadron colliders.

Another important issue is about the nonfactorization effects
of this process.
In the above, we use the general {\it suppression factor} ${\cal F}_S$
to describe these effects.
Further investigation about the nonfactorization effects
is needed but is beyond the scope of this work.
Nevertheness, we believe that the nonfactorization effects may not destroy
the main results obtained in our paper,
and the process in this paper may also be viewed as
a test of the factorization of the hard diffractive processes in hadron
collisions.

To sum up, in this paper, we give the formula for the diffractive $\jpsi$
production at hadron colliders in LLA QCD by using the two-gluon exchange
model.
We introduce the color-octet mechanism to realize the color-octet $c \bar c$
pair evolving into $\jpsi$ meson.
Though there may exist large theoretical uncertainties in our calculations,
our results show the importance of diffractive $\jpsi$ production at
hadron colliders to the study of small $x$ physics,
the property of diffractive process, the nature of the Pomeron and even
for the test of color-octet production mechanism of heavy quarkonium.
We hope the experimental measurement will be carried on in the
near future.

This work was supported in part by the National Natural Science Foundation
of China, the State Education Commission of China, and the State
Commission of Science and Technology of China.


\newpage
\vskip 10mm
\centerline{\bf \large Figure Captions}
\vskip 1cm
\noindent
Fig.1. Sketch diagram for the diffractive $J/\psi$ production at hadron colliders
in perturbative QCD. The black box represents the long distance process
for color-octet $c\bar c$ pair in ${}^3S_1$ state evolving into $\jpsi$.

\noindent
Fig.2. The lowest order perturbative QCD diagrams for $\jpsi$ production
at hadron colliders.

\noindent
Fig.3. The differential cross section $d\sigma/dt$ ($t=0$) at the Fermilab
Tevatron as a function of the lower bound of $x_1$ in the integral of
Eq.(\ref{e5}).

\begin{figure}
\hbox{
\psfig{file=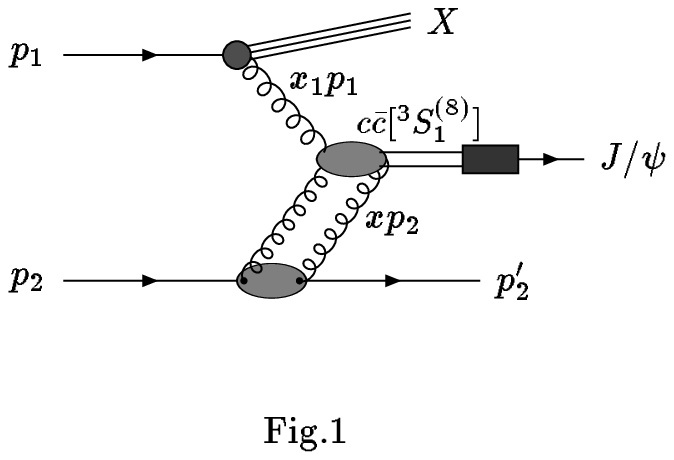,width=16cm,angle=0}}
\hbox{
\psfig{file=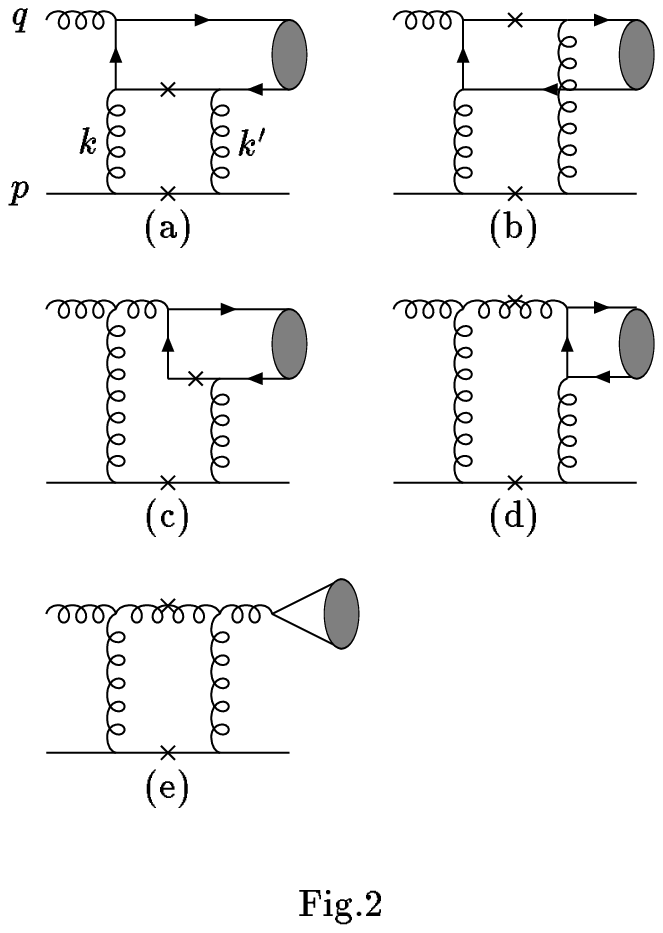,width=16cm,angle=0}}
\hbox{
\psfig{file=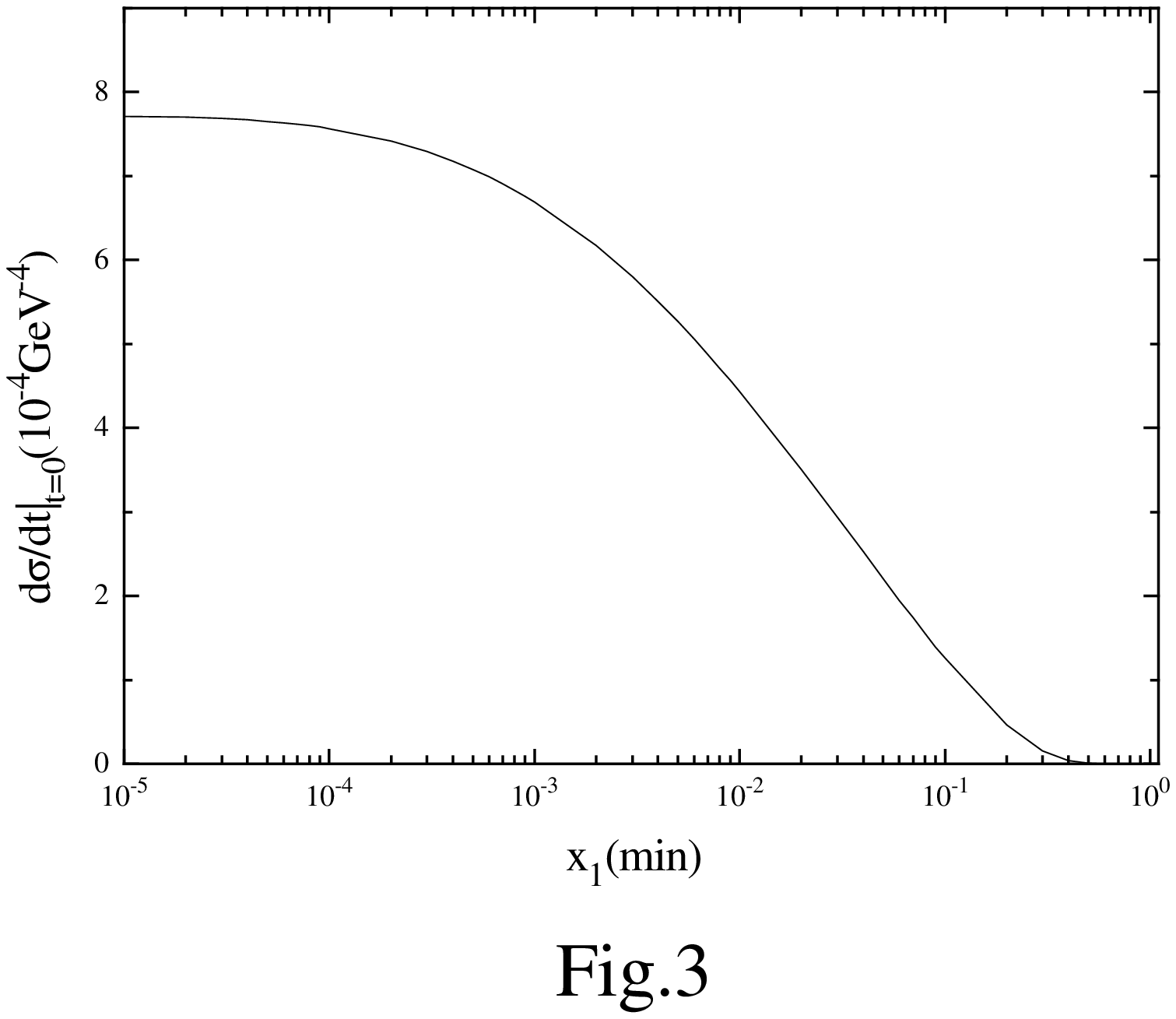,width=16cm,angle=0}
}
\end{figure}

\end{document}